# Levels of Binary Equivalence for the Comparison of Binaries from Alternative Builds

JENS DIETRICH, TIM WHITE, Victoria University of Wellington, New Zealand
BEHNAZ HASSANSHAHI, PADDY KRISHNAN, Oracle Labs Australia, Australia

In response to challenges in software supply chain security, several organisations have created infrastructures to independently build commodity open source projects and release the resulting binaries for Java/Maven and other software ecosystems. Build platform variability can strengthen security as it facilitates the detection of compromised build environments. Furthermore, by improving the security posture of the build platform and collecting provenance information during the build, the resulting artifacts can be used with greater trust. Such offerings are now available from Google, Oracle and RedHat. The availability of multiple binaries built from the same sources creates new challenges and opportunities, and raises questions such as: "Does build A confirm the integrity of build B?" or "Can build A reveal a compromised build B?". To answer such questions requires a notion of equivalence between binaries. We demonstrate that the obvious approach based on bitwise equality has significant shortcomings in practice, and that there is value in opting for alternative notions. We conceptualise this by introducing levels of equivalence, inspired by clone detection types.

We demonstrate the value of these new levels through several experiments. For this purpose, we construct a dataset consisting of Java binaries (jar files) built from the same sources independently by different providers, resulting in 14,156 pairs of binaries in total. We then compare the compiled class files in those jar files and find that for 3,750 pairs of jars (26.49%) there is at least one such file that is different, also forcing the jar files and their cryptographic hashes to be different. However, based on the new equivalence levels, we can still establish that many of them are practically equivalent; the number of pairs of jars with non-equivalent classes drops to 13.65 % in some cases. We evaluate several candidate equivalence relations on a semi-synthetic dataset that provides oracles consisting of pairs of binaries that either should be, or must not be equivalent.

CCS Concepts: • **Security and privacy** → **Software security engineering**; *Vulnerability management*; • **Software and its engineering** → *Software post-development issues.*

Additional Key Words and Phrases: Software Supply Chain Security, Build Security, Reproducible Builds

## 1 INTRODUCTION

Modern software is underpinned by complex and quickly changing software supply chains. Incidents and vulnerabilities like *solarwinds*, *codecov*, *equifax* and *log4shell* [21, 22, 36] have raised major concerns about the security of software supply chains [6]. This includes the security of complex and heavily automated processes used to build and deploy software. The possibility of attacks on these processes has long been hypothesised (the famous *Ken Thompson hack* [53]), but in recent years, such attacks have become common: examples include *xcodeghost*, *ccleaner*, *shadowpad*, *shadowhammer*, *solarwinds* and *xz* [7, 9, 36, 37].

A common approach to detect compromised builds of open source software is to create reproducible builds *(R-Bs)* [3, 32]. R-Bs enable independent parties to rebuild projects, and assess the resulting binaries for equality, usually by comparing cryptographic hashes. If two independently built binaries have the same hash, one can infer that the builds have not been compromised. This hinges on the assumption that an attacker is unlikely to be able to compromise both builds.

Recent empirical studies have confirmed that R-Bs are effective security measures [13, 22, 23], with building security identified as one of the main challenges in software supply chain security [22]. However, it has been also noted that "much of the software industry believes R-Bs are too far out of reach for most projects" [23]. Reproducing builds is still challenging and often fails [20, 30, 40, 49].

Authors' addresses: Jens Dietrich, Tim White, {jens.dietrich,tim.white}@vuw.ac.nz, Victoria University of Wellington, Wellington, New Zealand; Behnaz Hassanshahi, Paddy Krishnan, {behnaz.hassanshahi,paddy.krishnan}@oracle.com, Oracle Labs Australia, Brisbane, Australia.

A key challenge is to control build input. This can be addressed by jailing builds, i.e., controlling build input parameters in a highly sanitised environment. However, jailing has shortcomings such as build speed, limitations to control non-determinism and social restrictions [32]. Other security techniques like *ASLR* (Address Space Layout Randomization) fundamentally limit what can be achieved with input sanitisation [49]. As an alternative, explainable builds *(X-Bs)* have been proposed [17]. Instead of reproducing a build, X-Bs aim at explaining non-equivalences in the build artifacts that cannot be mitigated. Tools have emerged to facilitate this, including *BuildDiff* [35], *RepLoc* [41], *RepTrace* [42], *DiffoScope* [1] and *JarDiff* [5].

In recent years, several providers started to independently build and publish open source components. In particular, Google, Oracle and RedHat are now building and publishing Java components from open source at scale. Vendors claim additional quality attributes (such as SLSA-2 compliance, additional security analyses, and attestations), but do not position those as R-Bs of components built by developers and deployed on Maven Central [1]. Instead, the focus is to provide binaries that are more secure by design. However, this requires *substitutability*, i.e., that an alternatively built component can substitute an existing component in downstream applications. If the binaries are the same, then substitutability is guaranteed. However, if they differ, substitution becomes risky: it could break downstream clients due to behavioral changes or even suggest that one of the builds has been compromised. As we will demonstrate, this is rather common. Manually analyzing such situations is complex, error-prone, and labor-intensive. However, in many cases *equivalence*, sufficient to guarantee substitutability, can still be established by other means, such as comparing disassembled binaries.

We set out to study the equivalence of binaries for Java bytecode. While our main motivation is substitutability, having such equivalences has additional applications. By relaxing the postcondition of R-Bs from bit-for-bit identity to equivalence, R-B and X-B can be consolidated. Equivalence relations that incorporate provenance offer explainability when binaries differ, as required by X-Bs, while still preserving the key advantages of R-Bs: automated comparison of build results and the compact representation of those results through cryptographic hashes. This in turn makes it easier to achieve build security as more variability on input parameters can be tolerated, addressing some of the current limitations of R-Bs.

The paper is organised as follows. We first discuss the emergence of multiple alternative builds and their desirable quality attributes in Section 2. We present a dataset consisting of binaries built from the same sources by different providers, and set out to answer **RQ1: How common are binaries built from the same sources by alternative builds that are not bitwise equal?** A detailed analysis of why binaries (compiled classes) sometimes differ across alternative builds is presented in Section 3, answering **RQ2: What are the common reasons that classes are different across alternative builds?** We then discuss binary equivalence in Section 4, and introduce levels of binary equivalence in Section 5. This is followed by a discussion of binary equivalence relations based on readily available tools including decompilers, disassemblers and locality-sensitive hashes in Section 6. Using these binary equivalences, we answer **RQ3: How common are binaries built from the same sources by alternative builds that are not bitwise equal, but can be shown to be equivalent?** in Section 7. We continue to assess the various binary equivalence relations introduced earlier against the *BinEq* benchmark [19] in Section 8 in order to answer **RQ4: What is the correctness of binary equivalence relations that can be constructed from existing tools?** A discussion of related work (Section 9), and a short conclusion (Section 10) wrap up our contribution. We use a combination of engineering research, benchmarking and repository mining in this paper [39].

[1] https://central.sonatype.com/, retrieved 05/09/2024.

## 2 ALTERNATIVE BUILDS

### 2.1 The Emergence of Alternative Builds

Binaries built from open source are typically released into package repositories by developers. This process is often automated by build tools such as *Maven* or *Gradle*. Although it is considered best practice to build on a secure hosted build platform (a requirement for SLSA level-2 [4]), such as *GitHub* (using *GitHub Actions*), this remains optional from an open source developer's point of view, and many developers build locally.

Third parties can add value to open source by ensuring that builds are executed in a secure environment, thereby reducing the risk of compromised build environments, and facilitating provenance collection. For downstream users, this is attractive as it improves security and compliance (e.g., if SLSA compliance is required). A crucial aspect for this to work is that providers of alternative builds are able to reproduce the respective builds, i.e., locate the exact version of the source code that had been used, and construct a similar build environment (OS and compiler version, resource settings, etc.) [10, 24, 26, 30, 41]. In practice, this can be a difficult task, and may rely on the analysis of jar metadata, the assumption that sources are published alongside binaries in package repositories, the assumption that binary versions correspond to Git tags, metadata found in binaries, etc. This is complex, and likely to lead to variability; initiatives to improve this by publishing build specifications are emerging, such as *reproducible-central* [2].

### 2.2 Quality Attributes for Multiple Builds

Quality attributes for single builds are well understood. In particular, SLSA [4] defines such attributes. These relate to the security and trustworthiness of the build environment, and the generation of verifiable provenance. They are aggregated into levels (SLSA-L1 to SLSA-L3).

Given multiple builds, quality attributes of such a set are based on the attributes of the individual builds. However, there are additional qualities to consider. Additional attestations can be made about the relationships between multiple builds. In particular, *platform variability* matters.

Independent/hardened builds at scale will often choose to use a reference build as a baseline. In the Java/Maven open source ecosystem, this is usually the build used to produce the binary deployed on Maven Central, the default repository. This is a sensible choice as it maximises the chances of build success, which is often not easy to achieve [47, 52, 54]. However, even if an alternative build succeeds, this may not always result in attestations such as "build A confirms build B". For instance, consider the package `commons-io:commons-io:2.15.1`. The *sha1* hash of the binaries built and deployed by the developer on Maven Central is different from the *sha1* of the RedHat-built binary. There is an easy explanation here: there are differences in the metadata of both files, including in the manifests (*META-INF/MANIFEST.MF*), which have different `Build-Jdk-Spec` values [3].

Trying to closely reproduce a build environment will also make it more likely that a build is exposed to the same vulnerabilities that may have affected the build to be reproduced. Mutating the build environment on the other hand may reveal a *compromised build*. For instance, consider a build that builds a binary with a backdoor, injected by a compromised compiler. The backdoor consists of an additional call site to a network or system API in a function reachable from an entry point function (such as *main*). A second build using a different platform with an alternative compiler that has not been compromised would produce a binary that does not have such a call site, facilitating the detection of the backdoor, and therefore the compromised compiler.

The problem with this approach is that using a different build environment (including compiler) often leads to a different binary, even in the absence of any compromise. Compilers, and even compiler versions, differ in their optimisations, resulting in different binaries. In recent years there have been significant changes to how some of the most common Java code is compiled [46], such as string concatenation (JEP280) [50] and inner class member access (JEP181) [44].

[2]https://github.com/jvm-repo-rebuild/reproducible-central/, retrieved 05/09/2024

[3]See https://github.com/binaryeq/bineq-study-supplementary/tree/main/mvnc-vs-rh1/example10 for details.

Providers of alternative builds do not generally aim for full reproducibility, i.e., producing binaries identical to existing binaries in some repository. Instead, they focus on providing alternative binaries with quality attributes derived from a secure build environment that can be used as replacements of existing binaries. This however requires substitutability, i.e., a guarantee that the behaviour of the respective binaries does not change. If the binaries produced are identical, substitutability is guaranteed. The question arises what happens if this is not the case.

### 2.3 Providers of Alternative Builds

We first studied the consistency of binaries created and distributed within the Java/Maven ecosystem by four different providers: Maven Central (*mvnc* – the Maven default repository where binaries built by open source projects are deployed), Google Assured Open Source Software [4] (*gaoss*), RedHat [5] (*rh1* and *rh2*) and Oracle build-from-source (*obfs*). *Obfs* is an Oracle Inc. in-house repository containing 320 versioned artifacts (identified by a group, artifact and version id, i.e., GAVs) that we augmented with 1,712 GAVs from Oracle's public Maven repository [6].

A dataset to experiment on was constructed as follows. Starting from the 1,200 most-frequently used Maven unversioned artifacts (identified by group and artifact id, i.e., GAs) according to `libraries.io` [7], we removed 2 GAs which did not have Maven Central as their package URL. To increase overlap across providers, we then added the 699 GAs of the 812 GAs in *obfs* that were not already present in the `libraries.io` GA set, resulting in a baseline of 1,897 GAs to consider.

We then attempted to download binary and source jars for all versions (GAVs) of these GAs from each provider, filtering out those we determined to be invalid (due to missing binary or source jar, incorrect MD5 hashes, or jar extraction failures).

The remaining valid GAVs were canonicalised by stripping any provider-specific sub-patch-level version suffix (*-redhat-1*, *.redhat-00001*, etc). For RedHat, this resulted in cases where multiple artifacts correspond to a single artifact in Maven Central. For each GA a file *maven-metadata.xml* exists that lists those versions in the order of their release. We used this information to build two datasets for the RedHat repo, one using the first (*rh1*), and one using the last versions (*rh2*) available for each artifact. There are 977 normalised GAVs for which these two providers differ. Some *obfs* artifacts also have suffixes, but there were no collisions at the GAV level after canonicalisation. Since our interest is to compare the binaries built and distributed by different providers, and based on alternative builds, we first calculated the overlap between those repositories. We consider all GAVs present in at least two providers. The respective counts are shown in Table 1. The numbers in the diagonal represent the unique GAVs in (the downloaded part of) each repository. In the table caption we also report the total number of distinct pairs of jars from different providers, not counting the values in the diagonals. We analysed 16,621 pairs of jar files in total.

### 2.4 Source Equivalence

Our main interest is the variability of binaries obtained from alternative builds. Initially we assumed that if two artifacts exist for the same GAV, then they would have been built from the same sources. We first checked whether this was actually the case. Sources are packaged during the build and usually distributed alongside binaries. Maven Central has a requirement to supply sources [8] and distributes them alongside the binaries in a *-sources.jar* file; providers of alternative builds do the same. We used this information to compare source files first. We checked this for source code files (Java, Groovy, Kotlin, Clojure, Scala) that can be compiled into Java *.class* files. Surprisingly, there were multiple instances where the sources did not match.

[4] https://cloud.google.com/security/products/assured-open-source-software, retrieved 05/09/2024.
[5] https://maven.repository.redhat.com/ga/, retrieved 05/09/2024.
[6] https://maven.oracle.com/public/, retrieved 05/09/2024.
[7] https://libraries.io/api, retrieved 03/09/2024.

However, there are patterns where source may differ, but those differences do not cause behavioural changes in the respective binaries. These include changed comments, code re-formatting and the use of different character sets. We also found instances of another pattern related to annotations used for code generated during builds (a common practice in Java, often used to generate parsers for domain-specific languages from formal grammars). Some generators annotate generated code with the *javax.annotation@Generated* annotation and set its *date* property to the build timestamp, making the generated sources inherently non-deterministic. However, the annotation has a *RetentionPolicy.SOURCE* retention, so does not affect the resulting bytecode.

We took this into account by relaxing the requirement for *.java* source code files from strict string comparison to source *equivalence*, established by means of a custom AST-based comparison based on the *javaparser* library [9]. For sources of other JVM languages, strict string comparison is used. Table 2 shows, for each pair of providers, the number of artifacts (GAVs) for which binaries and sources are available and the sources are equivalent in this sense. There are 14,156 pairs of jars built from equivalent sources by different providers.

All experiments in this paper were run using the following environment unless stated otherwise: Apple M1 Pro, 32 GB memory running on MacOS Sonoma 14.5 with a Java(TM) SE Runtime Environment (build 17.0.2+8-LTS-86) Java HotSpot(TM) 64-Bit Server VM. Scripts implemented in Java used the following JVM arguments: *-Xmx20g -Djava.util.concurrent.ForkJoinPool=6* and *-Xss128m*.

Table 1. Number of artifacts provided by both provider1 and provider2. The total number of such pairs is 16,621.

| | mvnc | gaoss | rh1 | rh2 | obfs |
|---|---|---|---|---|---|
| mvnc | 8,109 | 1,509 | 4,095 | 4,095 | 2,032 |
| gaoss | | 1,514 | 269 | 269 | 84 |
| rh1 | | | 4,104 | 4,104 | 82 |
| rh2 | | | | 4,104 | 82 |
| obfs | | | | | 2,032 |

Table 2. Number of artifacts provided by both provider1 and provider2 where both artifacts are built from equivalent sources. The total number of such pairs is 14,156.

| | mvnc | gaoss | rh1 | rh2 | obfs |
|---|---|---|---|---|---|
| mvnc | 8,109 | 1,421 | 3,130 | 3,102 | 1,999 |
| gaoss | | 1,514 | 190 | 190 | 82 |
| rh1 | | | 4,104 | 3,905 | 68 |
| rh2 | | | | 4,104 | 69 |
| obfs | | | | | 2,032 |

```
1  public final class PackageVersion implements Versioned {
2      public final static Version VERSION = VersionUtil.parseVersion(
3 -        "2.17.2", "com.fasterxml.jackson.core", "jackson-core");
4 +        "2.17.2.redhat-00001", "com.fasterxml.jackson.core", "jackson-core");
```

Listing 1. Example source difference between *mvnc* and *rh1*

Listing 1 shows an example of non-equivalent sources between the *mvnc* and *rh1* providers of `com.fasterxml.jackson.core:jackson-core:2.17.2` [10]. The only difference is a version string in an automatically generated source file. We have also encountered examples where different sources were caused by the fact that the GAV version number did not match the Git tag [11]. This leads to non-matching sources if alternative providers locate the source code to build from using Git tags. The fact that the distributed sources in some cases differ is interesting, but not in scope of our study and omitted for space reasons. We therefore make source equivalence a precondition for pairs of artifacts to be studied further.

[8] https://central.sonatype.org/publish/requirements/, retrieved 05/09/2024

[9] https://javaparser.org/, retrieved 05/09/2024.

[10] See https://github.com/binaryeq/bineq-study-supplementary/tree/main/mvnc-vs-rh1/example6 for details.

Table 5. Number of artifacts with some .class files that are included in binary jars from both providers, but those files have non-equivalent contents. Strict bitwise equality comparison is used. The total number of such pairs is 3,750.

| | gaoss | rh1 | rh2 | obfs |
|---|---|---|---|---|
| mvnc | 400 | 1,490 | 1,467 | 45 |
| gaoss | - | 54 | 54 | 9 |
| rh1 | - | - | 214 | 8 |
| rh2 | - | - | - | 9 |

### 2.5 Bitwise Equality

Next, we looked into common compiled class files (*.class*) for each pair of binaries. For this purpose, we extracted the set of *.class* files from each jar, and compared these sets to collect the artifacts where some class files are contained in the jar built by provider 1, but missing in the jar built by provider 2. The results are shown in Table 3. Providers 1 and 2 are represented by rows and columns, respectively. Inspecting these reveals some common patterns: compiler-generated class files (class names containing $), classes in *META-INF/versions* corresponding to multi-version releases, and *package-info* classes. But even if those are ignored, there are still a few cases of different classes in binaries built by different providers, as shown in Table 4. This is however very rare (less than 0.2% (26/14,145) of the binaries compared). *org.codehaus.groovy:groovy:2.5.23* is an example of a project in which 580 class files present in the Maven Central jar are missing from the Google AOSS jar. [12]

Table 3. Number of artifacts with java bytecode code files (.class) in provider1 jar file missing from provider2 jar.

| | mvnc | gaoss | rh1 | rh2 | obfs |
|---|---|---|---|---|---|
| mvnc | - | 13 | 64 | 59 | 4 |
| gaoss | 6 | - | 5 | 5 | 1 |
| rh1 | 79 | 0 | - | 1 | 1 |
| rh2 | 78 | 1 | 6 | - | 2 |
| obfs | 7 | 1 | 3 | 3 | - |

Table 4. Number of artifacts with different java bytecode code files in provider1 jar file missing from provider2 jar, compiler-generated classes, multi-release classes and package-info classes are ignored.

| | mvnc | gaoss | rh1 | rh2 | obfs |
|---|---|---|---|---|---|
| mvnc | - | 2 | 9 | 9 | 0 |
| gaoss | 0 | - | 0 | 0 | 0 |
| rh1 | 3 | 0 | - | 0 | 0 |
| rh2 | 3 | 0 | 0 | - | 0 |
| obfs | 0 | 0 | 0 | 0 | - |

Then we analysed the content of the common *.class* files included in the jars of both providers being compared. Using strict bitwise comparison reveals that for a significant number of artifacts, there is at least one such *.class* file that is different. The results are shown in Table 5.

[11] See https://github.com/binaryeq/bineq-study-supplementary/tree/main/mvnc-vs-rh1/example8 for details.

[12] See https://github.com/binaryeq/bineq-study-supplementary/tree/main/mvnc-vs-gaoss/example7 for details. The gaoss jar also contains 409 pairs of duplicate files, of which **3 pairs have different CRCs**.

As an example of an artifact that contains bytecode-level differences depending on the build used, consider the versions of the artifact *com.squareup.okhttp:okhttp:2.7.5* provided by Google AOSS and Maven Central. The bytecode of the class *com.squareup.okhttp.Headers* (among others) is not identical; Listing 2 shows the top section of the diff for the disassembled (*javap -c -p*) classes. This shows that the differences are caused by a different numbering within the constant pool. [13]

```
1 8c8
2 <        1: invokespecial #7                 // Method java/lang/Object."<init>":()V
3 ---
4 >        1: invokespecial #2                 // Method java/lang/Object."<init>":()V
```

Listing 2. Partial diff of disassembled *com.squareup.okhttp.Headers.class* in jars provided by Google AOSS and Maven Central for *com.squareup.okhttp:okhttp:2.7.5*

This allows us to answer RQ1 as follows:

**RQ1: How common are binaries built from the same sources by alternative builds that are not bitwise equal?**
We compared 14,156 pairs of binaries from alternative builds, built from equivalent sources. We found that a significant number (3,750, i.e., 26.49%) of those pairs contain at least one *.class* file that is different, therefore causing the respective binaries (jars) to be different.

### 2.6 Threats to Validity

We have made the assumption that the sources published in the various repositories are the sources that were used to create the binaries. This is usually ensured by build tools packaging and distributing sources alongside binaries. In particular, the Maven command *source:jar* binds to the *package* lifecycle phase, which invokes the *generate-sources* phase [14]; *Gradle's withSourcesJar* plugin works similarly. In theory this mechanism could be compromised. We consider this highly unlikely, in particular when hosted builds are used (*gaoss*, *obfs*, *rh*, at least some of *mvnc*).

We only implemented source equivalence checks for Java sources, but used strict string comparison for non-Java sources. This may have slightly reduced the size of our dataset.

## 3 CAUSE ANALYSIS

### 3.1 Causes for Binaries being Different

Using the same dataset as before, we analyse reasons why *.class* files are different across alternative builds. We study 45,486,984 pairs of *.class* files for this purpose. For each pair, the respective class files have the same name in jars obtained from different repositories but built from equivalent sources. For this purpose, we identify categories from known differences in Java compilation, and patterns reported in related literature [46, 57]. The baseline category *Any* counts pairs of classes with any bitwise difference. We found 488,032 such pairs. *Version* compares the bytecode version of *.class* files, while *Superclass* and *Interfaces* look into the declared superclasses and interfaces, respectively. We further compare members (fields and methods), distinguishing between synthetic and non-synthetic methods and fields. The reason for this is that the synthesis of synthetic members is influenced by compiler-internal specifics. We also analyse the presence/absence of two pervasive changes in Java compilation strategy (*JEP181* and *JEP280*). Finally, we compare the constant pools of classes in general (i.e., content and order – *ConstPool*), and specifically for order (*ConstPoolOrder*). I.e., for a pair of classes to be reported in the *ConstPoolOrder* category, the respective constant pools must have the same entries, but in different order. The respective analyses are implemented using *ASM* [2], except for *Any* (a simple comparison of bytes is used), and the comparison of constant pools, which relies on analyzing the output of *javap -v* [15].

[13] See https://github.com/binaryeq/bineq-study-supplementary/tree/main/mvnc-vs-gaoss/example3 for details.
[14] https://maven.apache.org/plugins/maven-source-plugin/jar-mojo.html, retrieved 05/09/2024.

### 3.2 Results

Table 6 shows the results of these analyses. Many categories have a minimal impact. The bytecode version is a frequent cause, but it can also be controlled by specific build settings. However, the most common cause is discrepancies in the constant pools of classes. This is hardly surprising as many other changes (constants, call sites, supertypes, synthetic members etc) lead to constant pool changes. The most surprising finding is that many of the changes in the constant pool are only related to its order (see the example in Section 2.5).

In Table 7 we aggregate results for the jars containing these classes. In other words, a pair of jars will differ if they contain a pair of classes that vary due to this specific cause. These numbers are generally higher (except for *Version*, which usually applies to all classes within a jar), and now changes caused by synthetic members become more common.

Table 6. Causes of classes being different across alternative builds

| cause | count |
|---|---|
| Any | 488,032 (100%) |
| Version | 105,920 (21.7%) |
| Superclass | 25 (0.01%) |
| Interfaces | 91 (0.02%) |
| SynthMeths | 15,406 (3.16%) |
| NonSynthMeths | 883 (0.18%) |
| SynthFlds | 3,429 (0.7%) |
| NonSynthFlds | 165 (0.03%) |
| JEP181 | 932 (0.19%) |
| JEP280 | 606 (0.12%) |
| ConstPool | 394,877 (80.91%) |
| ConstPoolOrder | 255,687 (52.39%) |

Table 7. Causes of jars being different across alternative builds

| cause | count |
|---|---|
| Any | 3,750 (100%) |
| Version | 536 (14.29%) |
| Superclass | 3 (0.08%) |
| Interfaces | 16 (0.43%) |
| SynthMeths | 917 (24.45%) |
| NonSynthMeths | 41 (1.09%) |
| SynthFlds | 171 (4.56%) |
| NonSynthFlds | 32 (0.85%) |
| JEP181 | 7 (0.19%) |
| JEP280 | 9 (0.24%) |
| ConstPool | 3,142 (83.79%) |
| ConstPoolOrder | 2,128 (56.75%) |

A surprising result is that there are cases where superclasses differ across builds. We have inspected those examples, and found that they are all related to whether shading is being used. Shading is usually set up in the build configuration using build plugins that also support relocation, i.e., changing packages. Consider for instance the artifact `org.codehaus.groovy:groovy:2.5.23` and its class *org.codehaus.groovy.antlr.GroovySourceToken*. In the class built by *gaoss*, its superclass is *antlr/Token*, whereas in the class from the Maven Central artifact the superclass is *groovyjarjarantlr/Token*. This is caused by the *gaoss* build not correctly reproducing the shading part of the build that shades *antlr* and relocates its classes into the *groovyjarjarantlr* namespace: *groovyjarjarantlr.Token* vs *antlr.Token*. [16]

[15] We parse the constant pool from the output and resolve the constant pool addresses, retaining only the resolved values.
[16] See section "Superclass difference" in https://github.com/binaryeq/bineq-study-supplementary/tree/main/mvnc-vs-gaoss/example7 for details.

**RQ2: What are the common reasons that classes are different across alternative builds?**
Common causes for classes being different across alternative builds are general differences of constant pools (80.91%), constant pool order (52.39%), and different bytecode versions (21.7%). When considering differences that cause jars to be different, synthetic methods (24.45%) and fields (4.56%) are also significant contributors.

# 4 BINARY EQUIVALENCE

## 4.1 Build Equivalence

From an abstract point of view, a build can be modelled as a function that takes source code $s \in S$ as input and produces a binary, i.e.: $build : S \rightarrow B$. Two builds $build_1$ and $build_2$ are considered equivalent if they consistently produce the same binary from the same source code, i.e., $build_1(s) = build_2(s)$ for any $s \in S$. The resulting binaries can be substituted in applications, as binary equality implies behavioural equivalence.

In practice, we often observe some variation in those binaries. Missing build provenance often renders the creation of an alternative build function difficult. Binaries generated by different compilers or compiler versions can vary widely [44, 50], and compilers cannot be assumed to be deterministic [57] [17]. What is more, there are other sources of variability of builds, such as build time code generation. In Java, this feature is widely used to generate parsers from grammars or schemas, using build plugins for *antlr*, *javacc*, *jaxb*, etc. On top of this, compilation can be customised by annotation processing, used in popular frameworks like *lombok*, *spring* or *hibernate*. Finally, increasingly Java projects contain non-Java code (Kotlin, Groovy, etc) that is compiled into JVM bytecode through alternative compilers. Any of those tools may contribute to build variability or even non-determinism, but might not have an effect on the behaviour of the generated binaries. I.e., substitutability can still be achieved. We therefore relax the bitwise equality requirement to *binary equivalence*, i.e., $build_1(s) \simeq_{bin} build_2(s)$ for some suitable binary equivalence relation $\simeq_{bin} \subseteq B \times B$.

The results presented in Section 2.4 also suggest that sources also have some level of variability, such as timestamps inserted by build time code generators, inserted legal comments, formatting and char set issues, etc. Therefore, we also introduce the notion of equivalent sources using an equivalence relation $\simeq_{src} \subseteq S \times S$, and define *build equivalence* as follows:

*Definition 4.1 (Build Equivalence).* Two builds $build_1, build_2 : S \rightarrow B$ are called equivalent with respect to a source equivalence relation $\simeq_{src}$ and a binary equivalence relation $\simeq_{bin}$ iff $\forall s_1, s_2 \in S$ : $s_1 \simeq_s s_2 \Rightarrow build_1(s_1) \simeq_{bin} build_2(s_2)$.

## 4.2 Defining Binary Equivalence

Establishing whether two binaries $b_1$ and $b_2$ are equivalent is a matter of defining an equivalence relation $\simeq_{bin}$ between binaries, and then determining whether $b_1 \simeq_{bin} b_2$ or not. Such an equivalence is expected to be reflexive, symmetric and transitive. The perfect equivalence is behavioural (semantic) equivalence ($\cong$). Unfortunately, establishing behavioural equivalence is undecidable [31]. However, it is still possible and useful to *soundly under-approximate behavioural equivalence*, i.e., to devise equivalence relations such that $\simeq_{bin} \subseteq \cong$.

Java binaries are jar files, those jar files are archives consisting of metadata, resources, and finer-grained, atomic binaries representing compiled classes (*.class* files). Establishing the equivalence of jars can be reduced to the equivalence of their contents, using existing mechanisms to compare character-based metadata and resources, and some standard mechanism for other embedded binary resources such as media files. We therefore focus on the equivalence of *.class* files from here on.

[17] In Java, compiler non-determinism is seen as not desirable and if discovered is treated as a bug. Recent issues in OpenJDK's *javac* include: JDK-8264306, JDK-8072753, JDK-8076031 and JDK-8295024 (URL: *https://bugs.openjdk.org/browse/<issue-id>*).

### 4.3 Similarity-Based Binary Equivalence

One possible pragmatic approach to equivalence is to consider two binaries as equivalent if they are "similar enough". I.e., given a distance function $\Delta$ that measures how different two binaries are (similarity is the reverse of this), we could simply define

*Definition 4.2.* $b_1 \simeq_{bin} b_2$ iff $\Delta(b_1, b_2) \leq \tau$

for some given numerical threshold value $\tau$. Assuming that $\Delta$ is a well-behaving distance function [18], such a relation is reflexive and symmetric, but not transitive. This prevents certain types of inference like establishing equivalence via equivalent intermediaries.

Similarity-based methods are attractive as they can leverage a large body of research and tools developed mainly for the purpose of malware detection [8, 25]. Malware detection uses similarity, i.e., it detects binaries that are similar to known malware. When assessing binaries resulting from multiple builds, the approach is rather different: now differences (not similarities) may indicate the presence of vulnerabilities. This matters when considering the sensitivity to minor differences in binaries in both approaches, such as changed constant pool references or entries, or changed arithmetic instructions. A similarity function developed for malware detection will often tolerate such changes to some extent (i.e., still flag binaries as similar) as they may have been made by adversaries trying to obfuscate malware. On the other hand, when comparing binaries built from the same source, any differences are suspicious unless they can be attributed to features of the different build environments.

### 4.4 Transformation-Based Binary Equivalence

An alternative approach to check binaries for equivalence is to apply some normalising transformations to them, and then compare the results. If no information is lost during the transformation, such equivalences are inherently sensitive to even the smallest changes in behaviour. Given two binaries $b_1$ and $b_2$ and a transformation $t$ we define:

*Definition 4.3.* $b_1 \simeq_{bin} b_2$ iff $t(b_1) = t(b_2)$

I.e., the equivalence relation is the *equivalence kernel* of the transformation function. Such a transformation can be based on readily available tools like decompilers, disassemblers or cryptographic hash functions. Transformations map binaries into some other domain that is then used to establish equivalence, such as strings, numbers, or more complex data structures like ASTs.

An immediate advantage of defining equivalence via transformations is that those transformations can be composed (chained). For instance, a transformation to remove debugging symbols can be composed of several simple transformations, each removing one type of debug symbol. Another example is a transformation that decompiles a binary into source code, followed by another transformation that builds an AST and transforms or removes some nodes or branches.

Chaining generally reduces the complexity of the relation and the transformation defining it. The ability to compose transformations also facilitates the collection of provenance. In particular, standards diff tools can be applied to the transformed binaries in order to provide provenance. For instance, consider transformations producing human-readable output, such as the decompiler or disassembler-based transformations to be discussed in Section 6. Diffing those outputs with standard tools will reveal useful insights to the engineers who are assessing the builds.

[18] This means that the distance of an element to itself is zero, the distance value is always positive, symmetric in its arguments, and the triangle inequality holds.

The transformation-based approach facilitates the creation of hashes at the end of the transformation chain – hashes can simply be computed from the results of the transformation(s). For instance, assume that we decided to use a transformation based on a disassembler. Then the hashes of the disassembled binaries could be computed and published, facilitating the downstream verification of equivalence statements.

# 5 LEVELS OF BINARY EQUIVALENCE

## 5.1 Quality Attributes of Equivalence Relations

If binaries from multiple builds are available and can be used to support security-related decisions, then the quality attributes of the relations (i.e., the software embodying them) used to establish equivalence become relevant as a decision has to be made whether to trust equivalence statements or not. Those qualities include generic quality attributes that apply to any software [27], and attributes such as the maturity of a tool, the (trust in the) community or organisation providing a tool, technical attributes like error-proneness or flakiness, etc. In addition to this, it is desirable for equivalence relations to provide *provenance* for the (non- ) equivalence statement being inferred. In particular, this applies to non-equivalence statements that should trigger further investigation. Provenance facilitates this as it enables engineers to recognise false positives. In this context, a false positive could be a spurious non-equivalence statement (i.e., $b_1 \not\simeq_{bin} b_2$ for a pair of binaries $b_1$ and $b_2$) that would usually trigger further investigations. Using provenance, an engineer may recognise that a tool detects some difference in the binaries being compared, but concludes that this is not security-relevant. Provenance generation can often be achieved easily, for instance, by using standard diff tools on (textual representations of) transformed binaries. We can further assess equivalence relations according to the complexity of the algorithms used. This is similar to and inspired by code clone levels [12]. We refer to them as *levels of binary equivalence*.

## 5.2 Level 1

Binaries are considered *level 1 - equivalent* if they consist of the same sequence of bytes. Level 1 equivalence relations must support the generation of provenance for non-equivalent binaries. This is trivial and can be achieved by pointing to the positions in the byte sequences being compared where the sequences differ. Level 1 relations are transitive by definition.

This is the maximum level, and is particularly attractive as this can be quickly established and verified by applying secure hash functions to the inputs. There is no transformation function required here, although we may think about this in terms of applying the trivial identity transformation for the sake of conceptual clarity.

## 5.3 Level 2

Binaries are considered *level 2 - equivalent* if the only differences are that (1) bytes corresponding to instructions that have no semantic effect on code execution are ignored and/or (2) there is an isomorphic mapping between the two byte sequences (perhaps ignoring the effects of (1)). Level 2 equivalence relations must support the generation of provenance for non-equivalent binaries and must be transitive. I.e., it must be possible to infer the equivalence of two binaries via a common third binary already shown to be equivalent to both.

The first condition is similar to how non-code (such as blank lines and comments) is handled in code clone detection. It is essentially the elimination of noise. An example of this is to ignore bytecode only used for diagnostics, such as the *SourceDebugExtension* [33, 4.7.11] , the *LineNumberTable* [33, sect 4.7.12] and the *Deprecated* [33, sect 4.7.15] attributes in Java bytecode. Those attributes are typically only used by tools such as debuggers or compilers.

The main use case for the second condition is the reordering of elements, such as the order of methods (functions) or constant pool entries. The example discussed earlier illustrates how constant pool ordering can lead to non equality (Section 2.5, Listing 2), and level 2 - equivalence could still be established here. The isomorphic mapping requirement means that different instructions that have some effect on code execution cannot be identified, even if they have the same or very similar semantics. I.e., no mapping into some abstract model ("lifting") takes place here. This is similar to how reformatting and renaming of local variables is handled in source code clone detection.

One could argue that the order of methods (fields, annotations, etc) has an effect on the program semantics. There are some examples where programmers make assumptions about the ordering, leading to non-deterministic program behaviour such as flaky tests [48, 51, 59]. We consider those cases of API misuse – the API makes no guarantees about the order of elements in the array returned, and the actual order even differs across different runtimes (JVMs) for the same bytecode.

Considering that the intention of equivalence relations is to soundly under-approximate behavioural equivalence, this suggests that at level 2 we can only require *soundiness* (sound in the absence of dynamic language features), a common approach in many program analyses [34].

### 5.4 Level 3

Binaries are considered *level 3 - equivalent* if bytecode sequences representing *semantically equivalent* instructions are considered equivalent. This is usually achieved through some mapping transformation, but this mapping is not required to be an isomorphism, and some information might be lost during the process. Level 3 equivalence relations must support the generation of provenance for non-equivalent binaries and must be transitive.

The transformation functions used for level 3 equivalence may perform replacement operations, e.g., replace bytecode instructions, either single instructions or sequences. An example of such a replacement is to normalise semantically similar operations to push values onto the stack (`ldc`, `ldc_w`, various constant operations such as `iconst_0` , `iconst_1`, etc). More complex mappings may map pre- and post JEP280 bytecode generated for string concatenation [50] into some common representation. Some of the concrete operators to be discussed in Section 6 fall into this category. For instance, while Java byte code uses dedicated bytecode instructions to add small constant values onto the stack, a decompiler will normalise those. *Jnorm*/*jnorm2* maps the over 200 Java bytecode instructions to a much smaller set of 15 *Jimple* instructions, also addressing several more complex patterns [46]. A common theme here is that equivalence is established through abstraction, i.e., the transformations are mappings of bytecode into some model-based representation, ignoring some details deemed to be irrelevant while retaining the semantics of the program.

At level 3 there are more corner case scenarios emerging where semantic equivalence cannot be strictly guaranteed in the presence of dynamic programming techniques. For instance, consider changes made to the Java compiler in JEP 181 [44]. This changes the generation of synthetic methods that are accessible through the reflection API.

### 5.5 Level 4

Binaries are considered *level 4 - equivalent* if bytecode sequences representing semantically similar (but not necessarily equivalent) instructions are considered equivalent. The generation of provenance for non-equivalent binaries and transitivity are *not* required for level 4 equivalence relations. This is the weakest level. Such relations can leverage many existing methods based on binary similarity, often involving statistical or AI-based approaches. Such methods often lack the ability to provide provenance as many AI-based solutions are blackbox in nature.

Table 8. Properties of Binary Equivalence Levels

| level | 1 | 2 | 3 | 4 |
|---|---|---|---|---|
| sound | yes | no | no | no |
| soundy | yes | yes | yes | no |
| provenance generation | yes | yes | yes | no |
| transitivity | yes | yes | yes | no |
| complexity | low | high | high | high |

### 5.6 Summary and Discussion

Table 8 summarises the four levels and their main properties. Rows 2 and 3 indicate whether the under-approximation of behavioural equivalence is sound or at least soundy.

When equivalences between binaries are used in statements, claims or attestations about qualities of builds, there are two aspects to consider: (1) the value of the statement (2) the trust we have in this statement. Using two very similar environments provides relatively little value as argued above: if one platform was compromised, it is likely that the second build would have faithfully replicated the compromised environment and the issue would remain undetected. However, this statement can be made with a high level of certainty as the equivalence can be established at a high level, perhaps even level 1. On the other hand, if the equivalence of very different builds can be established, such an attestation has a higher value as it provides evidence that the platforms used have not been compromised. The argument is purely statistical: an adversary is unlikely to compromise multiple independent environments in the same way. But the binaries resulting from those builds are likely to be significantly different, and therefore establishing equivalence is more difficult and may rely on a lower level of equivalence − more complex, and therefore less trusted. We argue that there is a pareto front between those two attributes.

## 6 BINARY EQUIVALENCE RELATIONS

We discuss several equivalence relations we experimented with. They are all implementations based on existing transformations or similarities.

### 6.1 Same Sequence of Bytes

The baseline equivalence is *bitwise-equal.* Binaries are represented as byte arrays, and two binaries are compared by a byte-by-byte comparison of those arrays. This is a level 1 equivalence relation, easy to verify using commodity tools such as cryptographic hashes. The (non)equality of such hashes also provides provenance in a widely accepted format.

### 6.2 Disassembling

The Java Development Kit contains *javap* [19], a disassembler that converts Java byte code into a textual representation. The transformation-based *disassembled* equivalence is defined by the transformation of applying *javap -c -p <class-file>* on bytecode, and comparing the textual outputs. The *-c* flag is used to print out the disassembled code, the *-p* flag is used to include code from private members. Listing 2 shows the (partial) output of this transformation being applied to two *.class* files being diffed. As with other tools, the semantics of *javap* may depend on its version.

[19]https://docs.oracle.com/en/java/javase/17/docs/specs/man/javap.html

We use *javap* in Java 17 [20] for the experiments. *javap* has an additional *-v* flag which causes additional information, including the constant pool and line number and local variable information, to be output. By omitting this flag, the *disassembled* equivalence relation ignores differences in line numbers (which may have changed due to comments or whitespace) and in constant pool values that are not referenced within the class, achieving some useful level 2 normalisation. For example, although the *mvnc* and *obfs* versions of `jakarta.ws.rs:jakarta.ws.rs-api:3.1.0` have identical `module-info.java` source files, target the same JVM version (55, i.e., JDK11) and were built using JDK versions that differ only in the patch-level version, the resulting compiled `module-info.class` files differ – but only in two version strings, which are only output by *javap* if the `-v` flag is present. [21] Using the *disassembled* equivalence relation thus allows this insignificant difference to be disregarded.

## 6.3 Decompilation

Decompilers translate binaries back into source code [15]. The *decompiled* equivalence we evaluated is based on the *fernflower* compiler from JetBrains [22]. An interesting normalisation feature is that *fernflower* correctly decompiles string concatenation pre- and post JEP280 (using `StringBuilder` and `invokedynamic`, respectively). This implies that the transformation is not an isomorphism, i.e., information is lost. A particular weakness we have encountered is the long running time, the fact that it often fails, and is flaky; see Section 7.2.

Since decompilers combine inner classes with their outer classes into a single *.java*, they need to be decompiled together. The decompiled equivalence compares binaries by applying three rules in order: (1) if the binaries are bitwise equal, then they are equivalent (2) if inner classes are compared, comparison is delegated to the comparison of the top-level classes, (3) top-level classes are equivalent iff decompilation results in the same compilation unit (*.java* file).

## 6.4 Normalisation with *JNorm*

*JNorm* is a tool recently proposed by [46] based on the intermediate *jimple* bytecode representation of the *soot* [55] static analysis framework. While the tool was designed for similarity detection, it can also be used as a transformation to establish equivalence. *JNorm* normalises bytecode, i.e., it removes some compiler-specific differences. We consider two equivalences based on *JNorm* transformations: *jnorm* applies the tool without any additional options, whereas *jnorm2* applies aggressive normalisations using the parameters *"-o -n -s -a -p"*.

## 6.5 Locality-Sensitive Hashing (tlsh)

*Tlsh* [38] is a locality-sensitive hash that has been used in malware detection. This can be used to define a similarity-based, non-transitive equivalence using a similarity threshold.

We define two similarity-based equivalences based on the *tlsh* implementation by trendmicro [23], with thresholds of 10 (*tlsh10*) and 100 (*tlsh100*), respectively. Those thresholds were chosen based on the *tlsh* documentation [24].

## 6.6 Levels

Table 9 summarises the levels of the various relations. The soundness / soundiness properties are based on an assessment of the principles underpinning the tool used to provide the equivalence. We evaluate this later in Section 8 and find a few cases where those relations are actually unsound. We think however that those are engineering issues that could be addressed by modifying or extending the respective tool.

[20] The experiments presented later are conducted using *javap* in JDK version 17.0.2

[21] See https://github.com/binaryeq/bineq-study-supplementary/tree/main/mvnc-vs-obfs/example5 for more details.

[22] https://github.com/JetBrains/intellij-community/tree/master/plugins/java-decompiler/engine, commit *e9fbef8*

[23] https://github.com/trendmicro/tlsh, version 4.5.0, retrieved 05/09/24

[24] https://github.com/trendmicro/tlsh/blob/master/java/src/main/java/com/trendmicro/tlsh/Tlsh.java

Table 9. Levels of Binary Equivalence Relations

| relation | level | type |
|---|---|---|
| bitwise-equal | 1 | trivial transformation (identity) |
| disassembled | 2 | transformation |
| decompiled | 3 | transformation |
| jnorm, jnorm2 | 3 | transformation |
| tlsh10, tlsh100 | 4 | similarity |

## 7 EQUIVALENCE OF ALTERNATIVELY BUILT BINARIES

### 7.1 Binary Equivalence

In order to answer RQ3, we studied the performance of several equivalence relations constructed based on existing tools. Detailed results for *disassembled* and *jnorm2* can be found in Tables 10 and 11, respectively. We also evaluated the other equivalence relations discussed in Section 6, those details can be found in the supplementary material repository [25].

Table 10. Number of artifacts with some .class files that are included in binary jars from both providers, but those files have non-equivalent contents. Equivalence is based on disassembly. The total number of such pairs is 2,776.

| | gaoss | rh1 | rh2 | obfs |
|---|---|---|---|---|
| mvnc | 361 | 1,077 | 1,055 | 32 |
| gaoss | | 47 | 46 | 7 |
| rh1 | | | 134 | 8 |
| rh2 | | | | 9 |

Table 11. Number of artifacts with some .class files that are included in binary jars from both providers, but those files have non-equivalent contents. Equivalence is based on jnorm representation (using aggressive normalisation ("-o -n -s -a -p")). The total number of such pairs is 1,932.

| | gaoss | rh1 | rh2 | obfs |
|---|---|---|---|---|
| mvnc | 151 ↯ 131 | 555 ↯ 215 | 536 ↯ 207 | 24 ↯ 8 |
| gaoss | | 9 ↯ 8 | 8 ↯ 9 | 1 ↯ 3 |
| rh1 | | | 31 ↯ 30 | 1 ↯ 2 |
| rh2 | | | | 1 ↯ 2 |

Applying these relations generally shows further improvements, however, we encountered numerous cases where the respective transformations failed with errors for *jnorm*, *jnorm2* and *decompiled*. The number of errors where the comparison of a pair of artifacts failed for at least one *.class* file is reported in tables following the ↯ symbol [26]. In a security setting, it is sensible to classify those pairs as non-equivalent. Common errors encountered with *fernflower* include resource issues (both out-of-memory and stack overflow), known issues related to the lack of support of *fernflower* for multi-version releases [27]. The *jnorm* tool does not provide many details of why the analysis sometimes fails; error messages only indicate that the respective *jimple* files cannot be generated.

These results suggest that there are low-hanging fruit where using some commodity tools can significantly improve current practice over the status quo, and that it is likely that this can be pushed further with emerging (yet not yet mature) tools like *jnorm*. Note that even if these equivalences still classify many binaries as non-equivalent, they are already valuable for the following reasons:

[25] https://github.com/binaryeq/bineq-study-supplementary/tree/main/RQ3-details

[26] If for a given component there is a pair of classes that are non equivalent, and another pair of classes where the comparison leads to an error, then this component is counted in both categories, *non-equivalent* and *error*.

[27] https://youtrack.jetbrains.com/issue/IDEA-285079/Decompiler-Support-multi-release-Jars

firstly, none of these tools is built for the purpose of establishing binary equivalence. We are merely re-purposing existing tools build for debugging, comprehension, binary similarity analysis or other purposes. Purpose-built tools are likely to find more equivalences. Secondly, these equivalences can establish equivalence where bitwise comparison fails. If bitwise comparisons (and therefore the comparison of cryptographic hashes) fails, engineers have the difficult task to perform some analysis to establish whether one binary has been compromised, and whether a binary is a valid substitution. The above results suggest that by using equivalence this can often be avoided, or can at least support this decision.

```
1 25c25
2 < v4 = interfaceinvoke v1.<ch.qos.logback.access.spi.IAccessEvent: java.lang.Class getClass()
      >();
3 ---
4 > v4 = virtualinvoke v1.<java.lang.Object: java.lang.Class getClass()>();
```

Listing 3. Example Jimple difference between *mvnc* (left) and *obfs* (right) under *jnorm2*

Although *jnorm/jnorm2* are capable of “normalising away” many semantically uninteresting bytecode differences, it is not a panacea. As an example of a semantically equivalent bytecode difference that persists even through the aggressive *jnorm2* normalisation, consider `ch.qos.logback.access.net.AccessEventPreSerializationTransformer` in `ch.qos.logback:logback-access:1.3.11`, considering the binaries from *mvnc* and *obfs* [28]. Listing 3 shows the difference in the Jimple generated by *jnorm2* for these classes. This is caused by a change in the compiler in Java 18 [29] not (yet) supported by *jnorm*. Another example comes from the class `com.google.common.jimfs.Handler` in `com.google.jimfs:jimfs:1.2`. Despite the fact that the source code for this class in the *gaoss* and *mvnc* providers is identical, and both build and target versions are also identical (JDK11 and JDK7, respectively), we find different, though semantically equivalent, bytecode in the binary jars for concatenating 3 strings. As the `javap` diff in supplementary information [30] shows, the difference extends beyond mere constant pool reordering.

We summarise the results for all equivalence relations discussed in Section 6 in Figure 1.

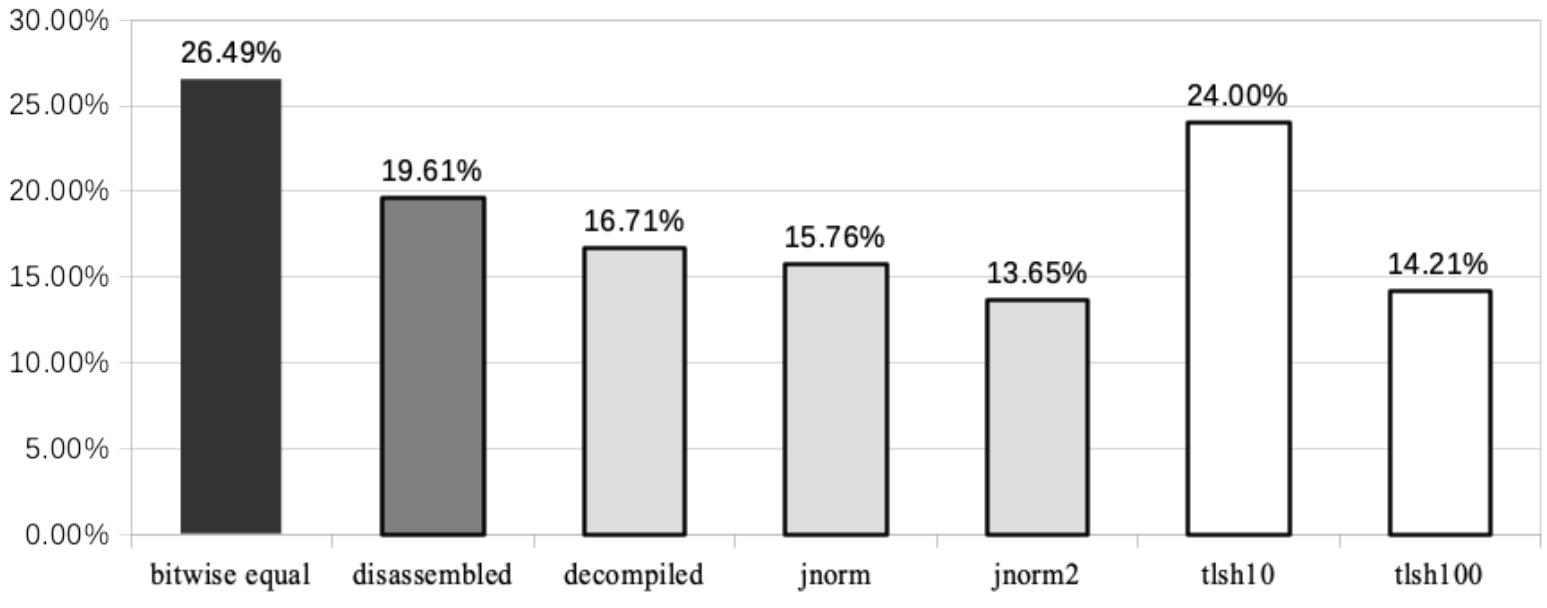

Fig. 1. Ratio of pairs of jars built by different providers that have at least one *.class* file that is not equivalent with respect to a given equivalence relation. The background colours of the bars indicate the level (1-4) of the respective equivalence relation as defined in Table 9.

[28] See https://github.com/binaryeq/bineq-study-supplementary/tree/main/mvnc-vs-obfs/example2 for further details.
[29] https://github.com/openjdk/jdk/pull/5165
[30] https://github.com/binaryeq/bineq-study-supplementary/tree/main/mvnc-vs-gaoss/example1

**RQ3: How common are binaries built from the same sources by alternative builds that are not bitwise equal, but can be shown to be equivalent?**
We compared 14,156 pairs of jars containing binaries from alternative builds, built from equivalent sources, with 26.49% of those pairs containing at least one *.class* file that is different (RQ1). By replacing equality by equivalence relations based on transformations provided by existing tools, this number can be significantly reduced to 19.61% (*disassembled*), 16.71% (*decompiled*), 15.76% (*jnorm*) and 13.65% (*jnorm2*), respectively.

### 7.2 Threats to Validity

We attempted to reproduce these results in a separate build environment (50GB Linux Mint 21.2 VM running on a 10-core 64GB i7-1355U 1.70 GHz Windows 10 machine), and found that the *fernflower* decompiler results exhibit some flakiness: The provider pair with the greatest difference was *rh2* vs. *mvnc*, which had 13 more inequivalent GAV pairs and 9 fewer crashing GAV pairs in the new environment. This amounts to $(13 + 9)/(850 + 72) = 2.3\%$ of all inequivalent-or-crashing GAVs for this provider pair. Overall, we found 2,362 non-equivalences or crashes across all provider pairs in this new environment, a difference of $(2366 - 2362)/2366 = 0.168\%$. We conclude that flakiness does not significantly impact the results reported for *fernflower*. The results for all other equivalence relations could be reproduced exactly.

The performance of the similarity-based relations (*tlsh10* and *tlsh100*, respectively) has to be taken with a grain of salt – it is easy to identify binaries as equivalent by crafting a relation that *over-approximates* semantic equivalence. For similarity-based methods, this can be achieved by increasing the similarity threshold. This is highly undesirable from a security point of view, as this might prevent the detection of compromised builds. The experiments presented in the next Section analyse this issue.

## 8 EXPERIMENTS 2 - BINEQ DATASET

### 8.1 Dataset

The *BinEq* dataset [19] consists of pairs (records) of compiled Java classes that either should be considered to be equivalent (equivalence oracles) or must not be considered to be equivalent (non-equivalence oracles). Equivalence records (*EQ*, 465,858 records) are obtained by building several open source projects with different compilers (versions, vendors and configurations), whereas non-equivalence oracles contains pairs of records with changes that affect the semantics of the bytecode, derived from API-breaking changes (*NEQ1*, 14,384 records), bytecode mutations (*NEQ2*, 141,585 records), and vulnerability patches (*NEQ3*, 202 records). In particular, the non-equivalence oracles enable us to test whether equivalence relations soundly under-approximate behavioural equivalence.

### 8.2 Metrics

For each combination of equivalence relation tested and oracle, we report the percentage of correct classifications. Correctness is defined as follows. Given a pair of classes equivalent accourding to an equivalence oracle $(b_1, b_2) \in EQ$, an equivalence relation $\simeq_{bin}$ classifies this pair correctly iff $b_1 \simeq_{bin} b_2$. Given a pair of classes non-equivalent according to a non-equivalence oracle $(b_1, b_2) \in NEQ$, $\simeq_{bin}$ classifies this pair correctly iff $b_1 \not\simeq_{bin} b_2$.

Table 12. Correct classification of pairs of classes

| oracle | EQ | EQ-SameMjCompVer | NEQ1 | NEQ2 | NEQ3 |
|---|---|---|---|---|---|
| disassembled | 0.545 | 0.919 | 0.996 | 1 | 1 |
| decompiled | 0.589 | 0.905 | 0.989 | - | 0.995 |
| jnorm | 0.749 | 0.91 | 0.988 | - | 1 |
| jnorm2 | 0.941 | 0.981 | 0.99 | - | 1 |
| tlsh10 | 0.235 | 0.784 | 0.97 | 1 | 0.985 |
| tlsh100 | 0.778 | 0.992 | 0.428 | 0.406 | 0.173 |

### 8.3 Results

The evaluation results can be found in Table 12. The correctness of the relations with respect to the equivalence oracle (EQ) shows that those relations can establish the equivalence of binaries that are not bitwise identical [31]. We report the results for EQ in general, and for the subset EQ-SameMjCompVer only containing records where the respective classes were compiled with the same compiler, and the same compiler major version. As the major compiler version is usually known when reproducing builds, those numbers are of interest. The performance of the various relations improves significantly when this assumption is made, reflecting the fact that the respective bytecodes are more similar when similar compilers are used.

We do not report correctness for *decompiled*, *jnorm* and *jnorm2* for NEQ2 due to the high error rates those tools encounter (over 80% of evaluations result in error). There are two reasons for this. Firstly, NEQ2 is a synthetic dataset, i.e., the byte code is synthesised by a mutation testing tool [19] . This bytecode may violate certain assumptions made by the respective tools about the bytecode. Secondly, NEQ2 construction injects changes by modifying individual *.class* files using pit mutations [32]. A common scenario is that such a mutation can be applied to some inner class, but not the outer class, or vice versa. This means that not all classes belonging to the same nest [44] are mutated. As the decompiler needs the entire nest to work, it will fail in situations like this.

The correctness of *tlsh100* with respect to the NEQ oracles is low as expected – similarity-based relations can be expected to be unsound, but by increasing the threshold for equivalence, the relation significantly over-approximates behavioural equality.

Of particular interest is the performance of the relations with respect to the NEQ oracles. While we cannot prove soundness (i.e., $\simeq_{bin} \subseteq \cong$, see Section 4.1) due to a lack of formal models underpinning those transformations, the non-equivalence oracles allow us to test this. There are a few cases where some of the relations report equivalence for NEQ records. I.e., those relations are technically unsound, even though they come close to soundness. As an example, consider *Base64TestData.class* (package omitted). There are two versions of this class in an NEQ1 record, one from *commons-codec-1.12-tests.jar* and one from *commons-codec-1.13-tests.jar*, respectively, both built with *openjdk-11.0.19*. This record is in the dataset because *revapi* correctly reports a semantic breaking change: The value of the constant `Base64TestData::CODEC_101_MULTIPLE_OF_3` was changed from *123* to *124*. [33]. However, disassembled does not show this change as unused constant values are omitted in the output of *javap -c -p*. [34] While such cases are rare (evidenced by the high percentages reported for the transformation-based relations in Table 12, columns NEQ1-NEQ3), they are a limitation of current tools (which were not build for this purpose). Better disassemblers could be build that avoid this particular issue.

[31] By design, *BinEq* equivalence oracles exclude pairs of classes that are bitwise equal.

[32] https://pitest.org/, retrieved 05/09/2024

[33] https://github.com/apache/commons-codec/commit/48b615756d1d770091ea3322eefc08011ee8b113, retrieved 05/09/2024

[34] See https://github.com/binaryeq/bineq-study-supplementary/tree/main/bineq/example9 for more details.

**RQ4: What is the correctness of binary equivalence relations that can be constructed from existing tools?** We evaluated the various equivalence relations introduced in Section 6 against the *BinEq* dataset, and found that the transformation-based relations can correctly classify between 54.5% and 94.1% of pairs of equivalent binaries. None of the relations soundly under-approximates behavioural equivalence, although all transformation-based relations come close, correctly classifying 98.8% or more non-equivalent pairs of binaries correctly. It is likely with some improvements of the tools underpinning those relations, soundness can be achieved.

### 8.4 Threats to Validity

There might be cases where the NEQ themselves are not correct, i.e., NEQ2 mutations that do not actually change the semantics. Such cases are theoretically possible. Addressing this would have led to slightly higher correctness values in the Table 12 NEQ columns. We note however, that those values are either 1 or very close to 1 already, and the examples discussed prove that there are cases of incorrect classification (i.e., unsound approximation of behavioural equivalence).

## 9 RELATED WORK

### 9.1 Comparing and Normalising of Binaries

Schott et al. propose *jnorm* [46], a tool to normalise Java bytecode based on *soot*'s *jimple* format [55]. This is an example of a transformation function that can be used to construct equivalence relations, and we discuss and evaluate this in this paper.

Xiong et al. study issues preventing reproducibility of Java-based systems [57]. They discuss differences in bytecode resulting from different builds, and develop a tool *JavaBEPFix* to remove certain types of build variability. Unlike *jnorm* and the other transformations discussed in Section 6, *JavaBEPFix* does not transform bytecode into some abstract model representation, but into (normalised) bytecode. This still fits within the framework of transformation-based equivalence we are using. Unfortunately, we were unable to obtain the tool to include in our evaluation, the authors informed us that this tool is not available to the public.

Numerous authors have worked on bytecode similarity [11, 14, 16, 28, 29, 45, 58]. This is related to our work as similarity can be used to define level-4 equivalence relations (although transitivity must be forfeited), as discussed is Section 4.3.

### 9.2 Reproducible Builds

Hassanshahi at al. [26] propose *macaron*, a toolkit for software supply chain security. They look into some of the challenges we have discussed that make it difficult to precisely reproduce builds, and may therefore lead to variability of binaries – the replication of the build environment, and the detection of the (exact version of the) sources used to build. *Macaron* supports Java/Maven. Keshani at al [30] also study how to locate the source code of (binary) Maven packages used to build from. This is not always straightforward, and our results obtained in Section 2.4 confirm this.

There is strong interest in the Linux community in reproducible package builds, and research supporting this. Ren et al. [41] and Bajaj et al. [10] study the prevalence and causes of unreproducible builds in Linux packages. Ren et al. [43] propose a tool *RepFix* to patch build scripts leading to non-reproducible builds, and evaluate this on Linux packages.

Goswami et al. [24] studies reproducibility of builds in *npm* and found that often builds were not reproducible. In their study, they successfully build 2,898 packages (versioned), of which 811 versions did not match. Differences can often be attributed to tools like *UglifyJS* [35] and *Babel* [36], and it is possible that equivalence could still have been established if suitable transformations and equivalence relations were used. For instance, such a transformation could be based on source maps created by *UglifyJS*. This suggests a situation similar to what we have observed in the Java/Maven ecosystem. A particular challenge is that unlike Java developers, JS developers often use dependency version ranges which are resolved at build time [18], adding to the variability of build results.

### 9.3 Double Compilation

Double-diverse compilation [56] uses a second compiler to safeguard the compilation of a compiler, therefore addressing the trusting-trust issue [53]. This is similar to the use of multiple alternative builds. In [56] Wheeler lists several inadaquate solutions, including *"A second compiler could compile the source code, and then the binaries could be compared automatically to argue semantic equivalence. There is some work in determining the semantic equivalence of two different binaries, but this is very difficult."* [56]. We argue that this can be generalised from compilation to more comprehensive build processes. And while we agree that semantic equivalence is difficult and in general even impossible to establish, we argue in this paper that under-approximating it (i.e., the detection of some but not all equivalent binaries) has merits.

## 10 CONCLUSION

We have demonstrated that binary equivalence is promising to assess the different binaries independently built from the same sources, and that this can facilitate the detection of compromised builds by removing false positives from the build analysis. We have presented a framework to classify such relations, and have evaluated how existing tools like *javap*, the *fernflower* decompiler, *jnorm* and *tlsh* can be re-purposed to build equivalence relations. It turns out that in many cases those tools can establish equivalence where binary equality and the comparison of hashes fail. However, more work is needed to construct equivalence relations that soundly under-approximate behavioural equivalence.

While we have made significant progress in studying binary equivalence for Java bytecode components, similar challenges exist in other ecosystems, especially in compiled languages with multiple compilers or rapidly evolving compilers, as well as in open and extensible build systems where plugins like code generators influence the outputs. Exploring binary equivalence in these contexts could offer valuable solutions for enhancing software supply chain security. Even for dynamic languages like JavaScript similar challenges exist as source code is manipulated in the build pipeline by applying code transformations (such as the TypeScript compiler and minifiers). However, further research is needed to fully investigate these possibilities.

## 11 DATA AVAILABILITY

The complete dataset used for RQ1–RQ3 is available from https://doi.org/10.5281/zenodo.14915249; the *BinEq* dataset used in RQ4 can be found in [19].

[35] https://www.npmjs.com/package/uglify-js, retrieved 05/09/24
[36] https://babeljs.io, retrieved 05/09/24